\documentclass[preprint]{revtex4}
\usepackage{enumerate}
\usepackage{graphicx}
\usepackage{dcolumn}
\usepackage{bm}
\usepackage{amsmath}
\usepackage{amsxtra}
\usepackage{amstext}
\usepackage{amssymb}
\usepackage{latexsym}
\usepackage{amsfonts}
\usepackage{amsmath}
\usepackage{amssymb}
\usepackage{graphicx}
\providecommand{\U}[1]{\protect\rule{.1in}{.1in}}

\begin{document}

\title{Analysis of the Thermonuclear Instability including Low-Power ICRH Minority Heating in IGNITOR}
\author{$^1$Alessandro CARDINALI\\$^1$ENEA for EUROfusion, Via E. Fermi 45, 00044 Frascati - Italy\\ Email: alessandro.cardinali@enea.it\\ and\\
Giorgio SONNINO$^{2,3}$\\$^2$Department of Theoretical Physics and Mathematics\\ Universit\'{e} Libre de Bruxelles (ULB), Campus Plain CP 231 \\Boulevard de Triomphe, 1050 Brussels, Belgium.\\ \&\\$^3$Royal Military School (RMS)\\Av. de la Renaissance 30 1000 Brussels - Belgium.\\ Email: gsonnino@ulb.ac.be\\
}

\begin{abstract}
The nonlinear thermal balance equation for classical plasma in a toroidal geometry is analytically and numerically investigated including ICRH power. The determination of the equilibrium temperature and the analysis of the stability of the solution are performed by solving the energy balance equation that includes the transport relations obtained by the classical kinetic theory. An estimation of the confinement time is also provided. We show that the ICRH heating in the IGNITOR experiment, among other applications, is expected to be used to trigger the thermonuclear instability. Here a scenario is considered where IGNITOR is led to operate in a slightly sub-critical regime by adding a small fraction of ${}^3He$  to the nominal $50$$\%$-$50$$\%$  Deuterium-Tritium mixture. The difference between power lost and alpha heating is compensated by additional ICRH heating, which should be able to increase the global plasma temperature via collisions between ${}^3He$ minority and the background $D-T$ ions.
\end{abstract}

\maketitle

\section{Introduction}\label{int}
Tokamak with a strong magnetic field like IGNITOR operates on the low-temperature branch of the ignition boundary \cite{coppi},\cite{freidberg} and \cite{mills} making impossible a stationary fusion reaction due to the thermonuclear instability. As a consequence of the instability, the self-heating of the plasma by alpha particles induces a significant rise of its temperature accompanied by an increase in the pressure, which in its turn will reinforce the thermal instability of the plasma. There has been a great effort, in the last decades, in investigating the various mechanism proposed for controlling the fusion thermal instability \cite{stacey}. In some work it was proposed that the balance in the growth of thermonuclear power be stabilized by increasing the energy losses from the plasma by changing the major radius $R_0$. Increasing $R_0$ ({\it i.e.}, adiabatic expansion) there will be a reduction of the plasma temperature. However there are serious engineering difficulties with this approach. Thus the possibility of significantly changing the large radius increases the volume of the chamber, which, obviously, will increase the volume of the magnet system \cite{putvinskii}. In other references it was suggested that $\alpha$-power could be regulated, by injecting pellets of fuel \cite{bespoludennov}. This method has significant advantages due to the technological progress of these last years in injecting a fuel tablet up to the center of the plasma column; this is connected with the fact that the tablet reaches a relatively high velocity \cite{migliori}. In addition, the required injection rate ($\sim 100 Hz$) is technically easy to achieve. The only difficulty remains that after the tablet is injected, the decrease in the cross section of the $D-T$ fusion reaction ($<\sigma v>\sim T^2$) will be compensated by the increase in the density, and the intensity of the thermonuclear reactions will remain unchanged. Control with modulation of the fueling rate and high-$Z$ impurity injection has also been demonstrated as an effective means for controlling the fusion thermal instability \cite{mandrakes}, especially when auxiliary power modulation cannot be used. The effects of a number of other phenomena on controlling the fusion thermal instability have been examined, and are: {\bf i)} transport losses due to toroidal magnetic field ripple via the $\tau_E$ term; {\bf ii)} impurity injection; {\bf iii)} the poloidal divertor; {\bf iv)} a soft beta limit; {\bf v)} compressing or decompressing the plasma; {\bf vi)} an ergodic magnetic limiter; {\bf vii)} modulation of divertor pumping; {\bf viii)} modification of alpha-particle transport; {\bf ix)} saw tooth oscillations; and {\bf x)} radial motion. A very exhaustive reference can be found in Ref.~\cite{stacey}. In other works it was proposed that the power of the thermonuclear burning be stabilized at a fixed level by regulation of the power of additional heating \cite{hartch}, \cite{kolesnichenko} and \cite{cardinali}. In particular in Ref.~\cite{cardinali} it was proposed that the reactor is operating in the sub-critical regime, {\it i.e.}, the parameters of the plasma are chosen so that the power of the thermonuclear reactions is slightly less than the power lost, for example by adding to the Deuterium-Tritium mixture a small fraction of ${}^3He$ (few percent); this small fraction of impurity unbalances the ideal ignition condition ($50\% - 50\%$ $D-T$), and the difference is compensated by additional heating. ICRH is, in fact, able to heat directly the minority species (ICRH minority heating) and by collision to transfer the power to the main species of the plasma: electrons and deuterium-tritium ions, by increasing the plasma temperature. The ICRH power acts to regulate the thermonuclear power via negative feedback. In this work this approach is accurately studied by solving the energy balance equation including the additional ICRH heating. Here the problems of ensuring stability of burning and the quality of transient processes for different confinement laws are studied. 

\noindent The main purpose of our work is to estimate the equilibrium temperature and the energy confinement time by assuming that the transport is governed by the classical kinetic theory. However, it is well known that there is almost always a strong anomalous diffusion in the outer plasma, for which a number of heuristic experimental scalings exist. So the classical thermal diffusion should be understood as an {\it ideal shape}, since the coefficient varies strongly with inverse temperature. Hence, we certainly do not claim to have estimated the confinement time in real experimental conditions, but only to show the results obtained by using the classical kinetic theory, applied to plasmas in toroidal geometry, without the aid of an {\it ad hoc} transport model. The next step, will be then to consider all the three collisional transport regimes ({\i.e.}, the  classical, Pfirsch-Schl$\ddot{\rm u}$ter and banana transport regimes) and to compare the analytical results with the solutions obtained by adopting a turbulent transport model like the gyro-Bohm model. This will be subject of future works. 

\noindent The manuscript is organized as follows. In section I the thermal energy balance equation in the framework of the neoclassical theory (see, for example, Ref.~\cite{balescu}) is recalled and all the terms of power gain and lost are specified. The energy-gain and -loss terms are evaluated in Sec.~\ref{sources}. Sec.~\ref{tl} is devoted to the determination and the discussion of the equilibrium solutions. The stationary thermal profile is determined by making use of the plasma dynamical equations and of the transport relations, which are rigorously obtained by kinetic theory. The estimation of the confinement time and the stability of the steady state thermal solution can be found in Sec.~\ref{ect} and in Sec.~\ref{ti}, respectively (a simplified calculation of the unstable modes is reported in the Appendix). The role of the ICRH power modulation in stabilizing/destabilizing the phenomenon is also herein discussed. Conclusions are reported in Sec.~\ref{conclusions}.

\section{One-dimensional toroidal plasmadynamical equations in the standard model and evaluation of the energy-gain and -loss terms}\label{sources}

\noindent Our first objective is to determine the electron temperature profiles of species $\iota$ ($\iota=e$ for electrons and $\iota=i$ for ions). This task will be accomplished by considering the balance equations (mass and energy) for species $\iota$ in toroidal geometry and by adopting the validity of the following standard model for the magnetic configuration
\begin{equation}\label{s1}
{\bf B}=B_0\Bigl(\frac{\epsilon\rho}{q(\rho)}{\bf e}_{\theta}+\frac{1}{1+\epsilon\rho}{\bf e}_{\phi}\Bigr)
\end{equation}
\noindent Here, ${\bf e}_{\phi}$ and ${\bf e}_{\theta}$ are the versors in the toroidal and poloidal directions, respectively, $\epsilon$ is the inverse of the aspect ratio: $\epsilon=a/R_0$ (with $a = 47cm$ and $R_0 = 132cm$ denoting the minor and the major radius of IGNITOR, respectively), and $B_0$ is the toroidal magnetic field at the magnetic axis (which in IGNITOR increases from $B_0 = 8T$ to $13T$ , during the ramp-up phase). $\rho$ and $q(\rho)$ denote the normalized minor radius ($\rho\equiv r/a$) and the safety factor, respectively. The safety factor profile we shall use for IGNITOR-plasma in this work is compatible with a plasma current of $11 MA$ at the end of the ramp-up phase and is greater then $1$ on the plasma magnetic axis. The equations of one-dimensional plasma dynamics, in toroidal geometry, by assuming the validity of the standard model, can be brought into the form (see, for example, Ref.~\cite{balescu})
\begin{eqnarray}\label{s2}
&&\!\!\!\!\!\!\!\!\!\!\!\!\!\!\!\!\!\!\!\!\!\!\!\!\!\!\!\!\!\frac{\partial n_e}{\partial t}=-\frac{1}{r}\frac{\partial }{\partial r}\Bigl(r<\gamma_r^e>\Bigr)\nonumber\\
&&\!\!\!\!\!\!\!\!\!\!\!\!\!\!\!\!\!\!\!\!\!\!\!\!\!\!\!\!\!\frac{3}{2}\frac{\partial p}{\partial t}+\frac{1}{r}\frac{\partial}{\partial r}\Bigl[r\Bigl(<q_e>+<q_i>+\frac{5}{2}(1+Z^{-1})T_e<\gamma_r^e>\Bigr)\Bigr]=\nonumber\\
&&\qquad\qquad\qquad\qquad\qquad\qquad\qquad\qquad\frac{c}{4\pi}\frac{E_0B_{0\phi}}{Rr}\frac{\partial}{\partial r}\Bigl(\frac{r^2}{q(r)}\Bigr)+S_{gain-loss}
\end{eqnarray}
\noindent where $p_e$ and $p_i$ are the plasma pressure due to the electrons and ions, respectively, and $n_e$, $T_e$ and $Z$ are electron density, electron temperature and the ion charge number, respectively. Here, $<\cdots>$ denotes the surface-average operation. $<q_\iota >$ and $<\gamma^e_r>$ are the averaged radial heat flux of species $\iota$ and the averaged electron flux, respectively. $E_0$ is the external electric field at $\rho=0$, and $S_{gain-loss}$ is the source term, {\it i.e.} the loss and energy-gain. Equation~(\ref{s2}) must be completed with the transport equations, {\it i.e.} with the thermodynamic flux force relations, in order to close the plasma dynamical equations. We make now several assumptions and approximations for reducing Eqs~(\ref{s2}) to a much simpler form. First, we assume that, in Eqs.~(\ref{s2}), the contributions related to the averaged electron flux $<\gamma^e_r>$ and the external electric field $E_0$, may be neglected with respect to the other terms. Second, the fuel is assumed to consist of a 50\%-50\% mixture of Deuterium ($D$) and Tritium ($T$), with a negligible concentration of $\alpha$-particles and ${}^3He$ (2-3\%). Third, the temperature of the plasma is the same for all species: $T_e = T_D = T_T=T $. Then Eqs ~(\ref{s2}), reduce to
\begin{equation}\label{s3}
\frac{\partial}{\partial t}\Bigl(\frac{3p_e}{2}+\sum_{i=D,T}\frac{3p_i}{2}\Bigr)+\frac{1}{r}\frac{\partial}{\partial r}\Bigl[r\bigl(<q_e>+<q_i>\bigr)\Bigr]=S_{gain-loss}
\end{equation}
\noindent From the local electro-neutrality condition we get
\begin{equation}\label{s4}
n_e=Z_Tn_T+Z_Dn_D=n
\end{equation}
\noindent and we have taken into account that $Z_D=Z_T=1$. In the calculation, we chose the following
profile for the electron density: $n_e(\rho)=n_e^a(1-\rho^2)+n_e^b$ , with $n_e^a =8.5\times 10^{14}(cm^{-3})$ and $n_e^b = 5\times10^{13} (cm^{-3})$, respectively. The total hydrodynamic pressure term is provided by the state equation
\begin{equation}\label{s5}
p=p_e+p_D+p_T=2nT
\end{equation}
\noindent and Eq.~(\ref{s3}) may be rewritten as
\begin{equation}\label{s6}
\frac{3}{2}\frac{\partial p}{\partial t}+\frac{1}{r}\frac{\partial}{\partial r}\Bigl[r\bigl(<q_e>+<q_D>+<q_T>\bigr)\Bigr]=S_{gain-loss}\end{equation}
\noindent Before discussing on the structure of the heat flux term (to which we will dedicate Sec. III), we determine the structure of the loss-gain terms on the r.h.s. of Eq.~(\ref{s6}). Note that in our equations we are assuming the physical quantities expressed in $cgs$ units (unless differently specified) so that the pressure is given in terms of $[m][l]^{-1}[t]^{-2}$, and the heat flux $[M][t]^{-3}$, in this manner Eq.~(\ref{s6}) has the dimension of a power density, and it represents the power density balance. The term on the r.h.s. of Eq.~(\ref{s6}) (the power gain-loss term) is specified as follows:
\begin{equation}\label{s7}
S_{gain-loss}=Q_\alpha+Q_b+Q_{add}
\end{equation}
\noindent where $Q_\alpha$ is the alpha heating power, $Q_b$ is the radiation loss (Bremsstrahlung), and $Q_{add}$ is the additional ICRH heating respectively. The alpha heating power density is given by the following formula
\begin{equation}\label{s8}
Q_\alpha=\frac{n^2}{4}<\sigma v>_{D-T}E_\alpha
\end{equation}
\noindent where $E_\alpha$ is the energy at which the alpha particles are created ($3.5 MeV$), $\sigma$ is the reaction cross section and it is a measure of the probability of a fusion reaction as a function of the
relative velocity of the two reactant nuclei, given in barn $[l]^{-2}$. If the reactants have a distribution of velocities, {\it e.g.} a thermal distribution with thermonuclear fusion, then it is useful to perform an average over the distributions of the product of cross section and velocity {\it i.e.} $<\sigma v>_{D-T}$ in units $[l]^3[t]^{-1}$. The reaction rate (fusions per volume per time) is $<\sigma v>$ times the product of the reactant number densities, $(1+\delta_{ij})^{-1}n_in_j<\sigma v>_{D-T}$, if a species of nuclei (deuterium) is reacting with another species (tritium), such as the $D-T$ reaction at $50\%$, then the product $n_in_j$ must be replaced by $\frac{n^2}{4}<\sigma v>_{D-T}$, which increases from virtually zero at room
temperatures up to meaningful magnitudes at temperatures of $10-100\ keV$ . At these temperatures, well above typical ionization energies ($13.6\ eV$ in the hydrogen case), the fusion reactants exist in a plasma state. In our calculation we have assumed that the dependence of the cross-section $\sigma $ on temperature is given by an analytical 3-parameter fitting \cite{sing},
\begin{equation}\label{s9}
\sigma(E)=\frac{\pi}{(2\mu/\hbar)E(m_b/m_a+m_b)}\frac{1}{\theta^2}\times\frac{(-4C_3)}{(C_1+C_2/E)^2+(C_3-1/\theta^2)^2}
\end{equation}
\noindent where $E$ is the energy of incident particles in the laboratory system,
\begin{equation}\label{snon1}
\theta^2=\frac{1}{2\pi}\Bigl[\exp\Bigl[\frac{2\pi}{\sqrt{2\mu E/\hbar^2}\ [\hbar^2/(\mu Z_iZ_je^2)]}\Bigr]-1\Bigr]\nonumber
\end{equation}
\noindent is the Gamow penetration factor, $\mu$ is the reduced mass, $\hbar$ is the Planck constant, and the fitting points for the 3-parameter fit formula in Deuterium Tritium reaction are: $C_1 = 0.5405$; $C_2 = 0.005546$; $C_3 = 0.3909$. By using Eq.~(\ref{s9}) we can estimate $<\sigma v>$ in terms of the plasma temperature. The calculation of $<\sigma v>_{D-T}$ can be performed by averaging the cross section over the relative velocities of the reactants keeping the relative ion energy distribution in plasma to be Maxwellian
\begin{equation}\label{snon2}
<\sigma v>_{D-T}=\frac{1}{m}\int_0^\infty\sigma(E)\exp\Bigl({-\frac{E}{K_BT}}\Bigr)dE\nonumber
\end{equation}
\noindent where $m$ is the mass particle, and $K_B$ the Boltzmann constant. 

\begin{figure}\resizebox{0.7\textwidth}{!}{%
\includegraphics{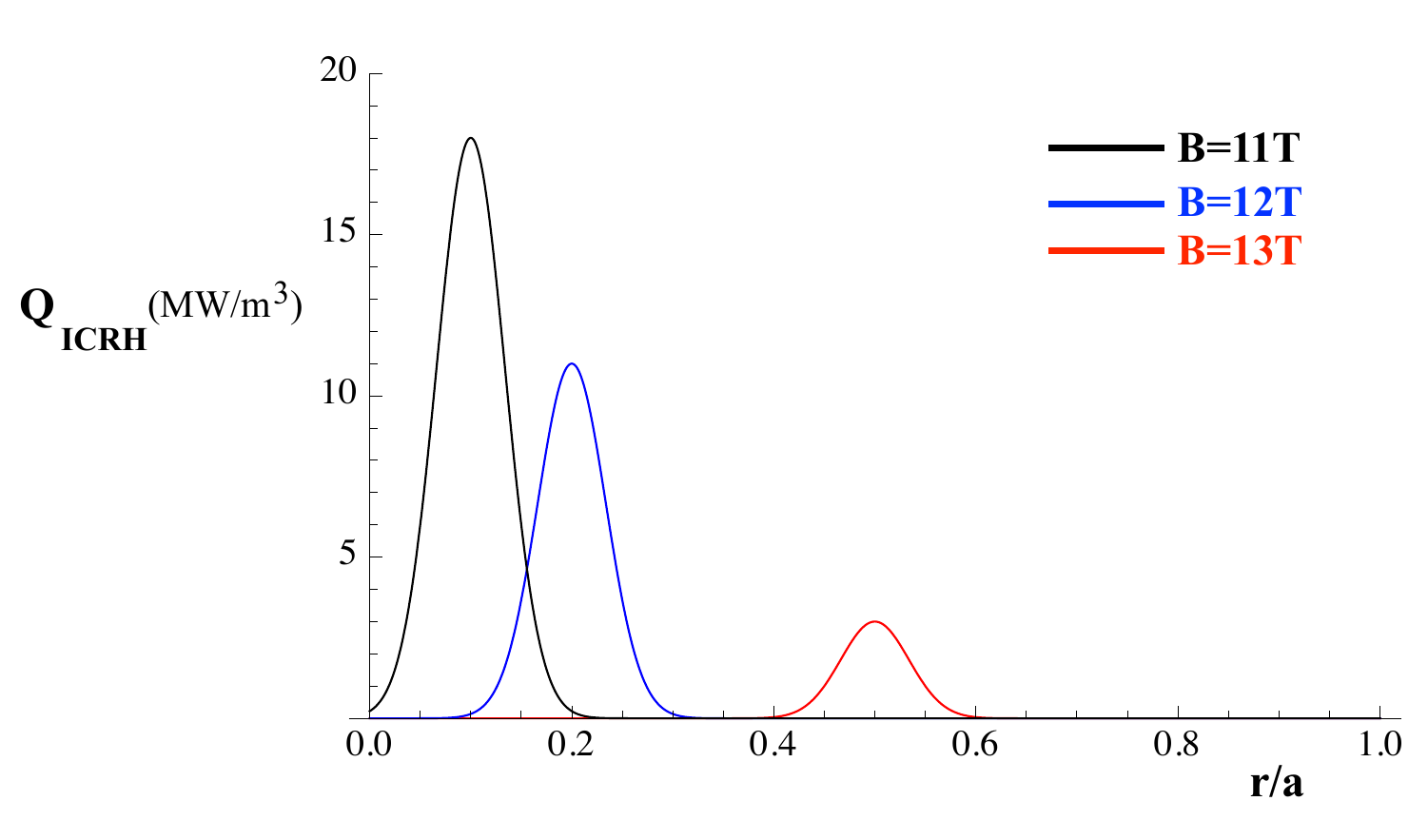}}
\hspace{1.5cm}
\caption{ \label{fig1} 
{\it RF power deposition (on the minority} ${}^3He$ (2\%)) {\it in} $MW/m^3$ {\it vs} $r/a$ {\it when the magnetic field is ramping up from} $11$ {\it (black line) to} $13$ {\it (red line)} {\it Tesla for the Ignitor plasma parameters in the ignited scenario. The applied frequency is} $115MHz$.
}
\end{figure}

\noindent The term $Q_b$ is the radiation loss (Bremsstrahlung)
\begin{equation}\label{s10}
Q_b=-1.69\times 10^{-25}n^2(cm^{-3})Z_{eff}T^{1/2}(eV)
\end{equation}
\noindent This formula gives directly the power density loss for bremsstrahlung in $cgs$ units with the temperature is measured in $eV$ and the number density in $cm^{-3}$. Finally the term due to the ICRH power absorption corresponds to $Q_{add} =Q_{ICRH}(r)$. To this end it is useful to recall that
the power deposition profile in IGNITOR ignited scenario with ramping magnetic field [from $8$ to $13T$] has been analyzed in detail in Ref. \cite{cardinali1} by using the ICRH full wave code TORIC \cite{brambilla}, coupled to the quasi-linear Fokker-Planck routine SSFPQL \cite{brambilla1}-\cite{brambilla2}. The best ICRH scenario to achieve an efficient absorption rate is the minority heating: in the case of IGNITOR, when a small fraction of ${}^3He$ (2-3\%) is added to the $D-T$ mixture, the first pass absorption on the ions near the center of the plasma column is very efficient. The remaining coupled power is damped on the electrons over a broad radial interval of the plasma column. The fundamental harmonic of ${}^3He$ is located in a radial interval $-0.5< r/a <+0.5$ (with $a$ denoting the minor radius of the tokamak) when the magnetic field is varied from $9$ to $13T$ and the antenna frequency is $f=115\ MHz$. For example, calculations of the power absorption level for three different external magnetic field ($11$, $12$ and $13T$), corresponding to three different times of the discharge evolution, for a plasma formed by D (50\%), T (50\%) and a small fraction of ${}^3He$ ($\simeq 2-3\%$), density and temperature between $5$ and $9\times10^{20}\ m^{-3}$, and from $4$ to $6\ KeV$ respectively, show that the peak of absorption is located at the fundamental harmonic of  ${}^3He$ and second harmonic of Tritium. In Fig. 3 the RF power deposition for a coupled power of $1.5\ MW$ is shown vs $\rho$ when the magnetic field is ramping up from $11$ to $13T$. It is possible to observe that the deposition is mainly concentrated at the fundamental harmonic of ${}^3He$, a small fraction being given to the electrons via Landau damping depending on the minority fraction; moreover, when the field increases, the resonance layer moves towards the periphery, but still remaining in the bulk of the plasma at $13T$. The power absorbed by the ${}^3He$ (minority heating) is quasi- linearly redistributed on the collisional time essentially to the Deuterium and Tritium bulk ions, with a fraction to the electrons. The consequence is that the plasma temperature increases accelerating the attainment of ignition.
\noindent An analytical expression for the power profiles inside the plasma can be deduced by fitting the numerical results giving a $Q_{add}=Q_{ICRH}(r)$ that is essentially independent on the bulk temperature
\begin{equation}\label{s11}
Q_{add}=\beta\exp\Bigl[\alpha\frac{B(\rho_{ICRH})}{B_0}\Bigr]\exp\Bigl[-\frac{(\rho-\rho_{ICRH})^2}{\Delta}\Bigr]
\end{equation}
\noindent The expression $B/B_0$ may be estimated by adopting the validity of the standard model. We get [see Eq.~(\ref{s1})] $B(\rho_{ICRH})/B_0\simeq1/(1+\epsilon\rho_{ICRH})$. The expression in Eq.~(\ref{s11}) fits very well the numerical curve (obtained by running TORIC+SSQLFP) by setting $\alpha= 15.3478$, $\beta= 6.59126\times10^{-6}MW / m^3$ and $\Delta= 0.0477032$ [see Fig~(\ref{fig1})].
\noindent To test the analytical fitting we integrate Eq.~(\ref{s11}) at $B=13T$ ($\rho_{ICRH}=0.5$), over the entire volume occupied by the plasma, and we obtain the total ICRH power, $P_{ICRH}$, injected into the plasma
\begin{equation}\label{s12}
\!P_{add}\!=\!P_{ICRH}\!=\!\int\! dV Q_{add}\!=\!\beta\int\! dV\exp\Bigl[\alpha\frac{B(\rho_{ICRH})}{B_0}\Bigr]\exp\Bigl[-\frac{(\rho-\rho_{ICRH})^2}{\Delta}\Bigr]\!\simeq\!1.5MW
\end{equation}
\noindent which coincides with the power input in the numerical code TORIC.

\section{Evaluation of the thermal loss}\label{tl}

\noindent The energy balance equation (6) should be completed with the transport equations relating the averaged thermodynamic flows $<q_e >$ and $<q_i >$ with the thermodynamic forces $-T_e^{-1}\partial_r T_e$ and $-T_i^{-1}\partial_r T_i$. The complete transport relations are composed by the sum of three terms: the classical, the Pfirsch-Schl${\ddot {\rm u}}$ter and the banana contributions. However in this work, at the first and simplest approximation, we shall study the case where the closure equations are provided solely by the classical term, appropriately estimated for a plasma in a toroidal geometry. The general situation, where all the transport contributions are taken into account, will be subject of a future work. Under this approximation, by kinetic theory we find that, for a plasma in toroidal geometry, the averaged total heat flow is related to the temperature gradient by the following equation \cite{balescu}
\begin{equation}\label{tl1}
<q_{tot}>=-\sum_{\iota=i,i}\Bigl(<\kappa_r^e(T)>_{CL}+<\kappa_r^i(T)>_{CL}\Bigr)\frac{\partial T}{\partial r}
\end{equation}
\noindent where we have taken into account that $T_i^{-1}d_rT_i=T_e^{-1}d_rT_e$. The expression for the electron and ion thermal conductivities $<\kappa_r^\iota>$ ( $\iota=e,i$) can be brought into the form
\begin{equation}\label{tl2}
<\kappa_r^\iota>_{CL}=\frac{5}{2}\frac{n_\iota T_\iota}{m_\iota}<{\tilde\kappa}_r^\iota>_{CL}
\end{equation}
\noindent where $\tau_\iota$ and $<{\tilde\kappa}_r^\iota$ are the collision time and the dimensionless averaged thermal conductivity of the species $\iota$, respectively. By kinetic theory we know that, {\it in toroidal geometry, the classical contribution to the transport coefficients coincides exactly with the asymptotic limit of the perpendicular transport coefficients estimated by the classical theory, averaged over a magnetic surface} \cite{balescu}. {\it For asymptotic limit we mean the value of the classical transport coefficients, estimated for} $\Omega_\iota\tau_\iota>>1$ , with $\Omega\iota$ and $\tau_\iota$ denoting the Larmor frequency and the collision time of species $\iota$, respectively. In other terms,
\begin{equation}\label{tl3}
\!<\kappa_r^e>_{CL}=c_{33}^e<\frac{1}{(\Omega_e\tau_e)^2}>\ \ ;\ \ <\kappa_r^i>_{CL}=c_{33}^i<\frac{1}{(\Omega_i\tau_i)^2}>=\Bigl(\frac{m_i}{m_e}\Bigr)^{1/2}\frac{c_{33}^i}{c_{33}^e}<{\tilde\kappa}_r^e>_{CL}
\end{equation}
\noindent Here, $c_{33}^\iota$ are the (dimensionless) coefficients of the linear collision matrix of species $\iota$, {\it i.e.}, $c_e =(13+4 2)/10$ (with $Z=1$) and $c_i =2\sqrt{2}/3$. Index $i$ in Eq.~(\ref{tl3}) stands for the {\it effective ion} with mass $m_i = (m_D + m_T)/2$ (since we assumed that the fuel is composed by $50\%$ of Deuterium and $50\%$ of Tritium) and $Z=1$. Note that, Eq.~(\ref{tl3}) takes into account the {\it toroidal geometry} of the Tokamak, but {\it not} the {\it inhomogeneity} and {\it curvature} of the magnetic field. As known, the latter is matter of the neoclassical theory. Moreover, in Eqs~(\ref{tl2}) and (\ref{tl3}), we have taken into account that the nonlinear corrections to the linear classical transport coefficients may be neglected \cite{sonnino1}, \cite{sonnino2} and \cite{sonnino3}.  At the steady state, we get
\begin{equation}\label{tl4}
-\frac{1}{r}\frac{d}{dr}r\Bigl(<\kappa_{tot}(T)>_{CL}\frac{d}{dr}T\Bigr)=Q_\alpha+Q_b+Q_{add}
\end{equation}
\noindent where $<\kappa_{tot}(T)>_{CL}=<\kappa_{\perp\infty}^e>+<\kappa_{\perp\infty}^i>$, and $Q_\alpha$, $Q_b$ and $Q_{add}$ are given by Eqs.~({\ref{s8}), ({\ref{s10}) and (\ref{s11}), respectively. Equation~(16) is the simplest version of the steady state power balance where the terms corresponding to the loss of energy density due to the expansion of the fluid and to the loss of energy density due to diffusive processes are neglected. These simplifications are justified from the fact that a magnetic fusion reactor is (almost) a steady state system with small and negligible flows. In addition, we are dealing with strongly magnetized plasmas and the turbulent effects are, therefore, notably reduced. Hence, at the first approximation, the time derivative as well as the convection and compression terms may reasonable be neglected. Fig. (\ref{fig2}) shows the total thermal conductivity (sum of the electron and the reduced ion thermal conductivities). In Figs~(\ref{fig3}) and (\ref{fig4}) the total source-density are reported, against the minor radius and temperature, in absence ($P_{ICRH}\simeq 0$) and in presence of additional source ($P_{ICRH}\simeq 1.5MW$), respectively.

\begin{figure}\resizebox{0.45\textwidth}{!}{%
\includegraphics{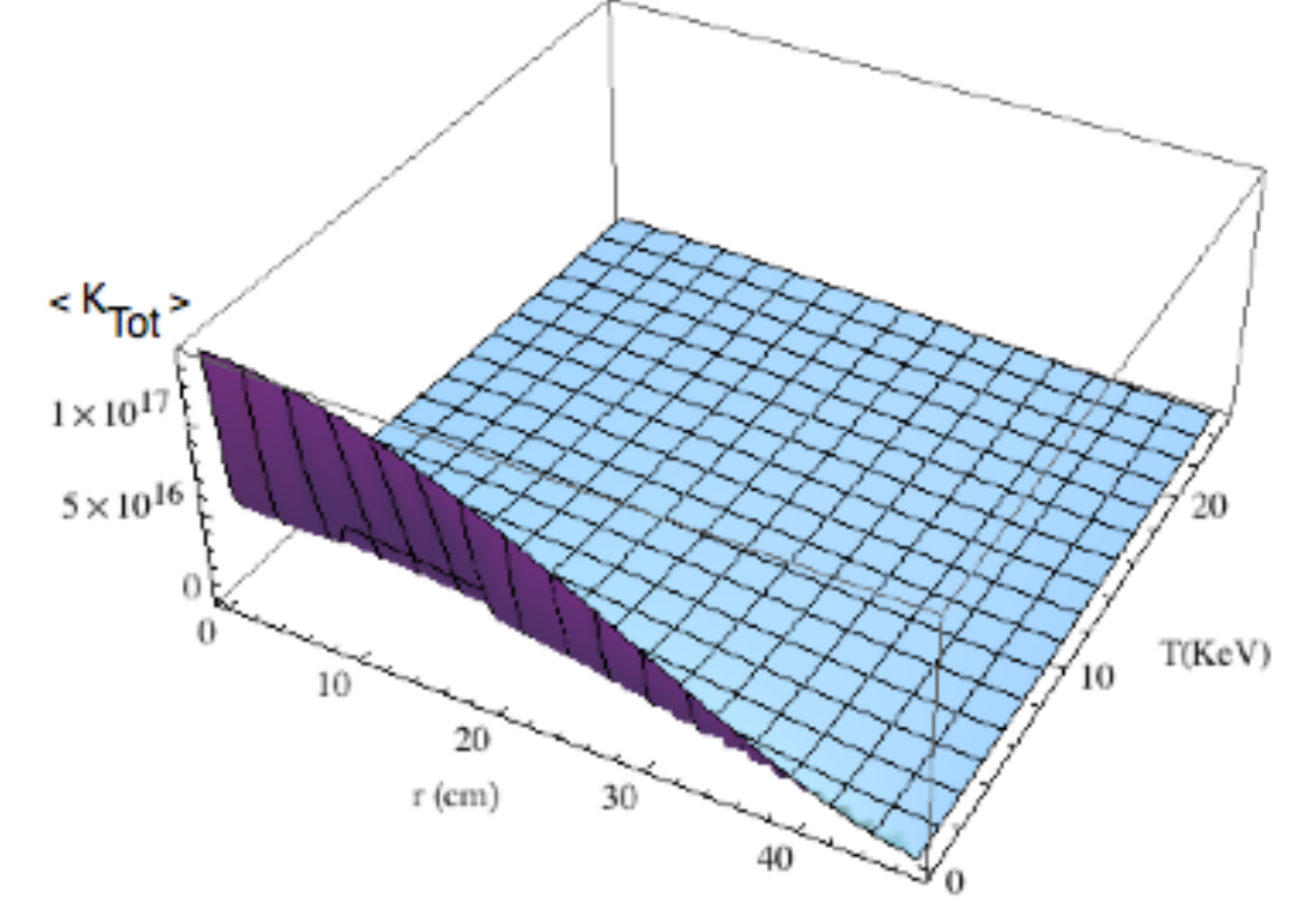}}
\caption{ \label{fig2} 
{\it Surface-averaged, total classical thermal conductivity coefficient for IGNITOR plasmas}, $<\kappa_{tot}>$ ($cm^{-1}sec^{-1}$) {\it vs the minor radius and temperature}.
}
\end{figure}
\begin{figure*}
\hfill
  \begin{minipage}[t]{.4\textwidth}
    \begin{center}  
\hspace{-3cm}
\resizebox{1.2\textwidth}{!}{%
\includegraphics{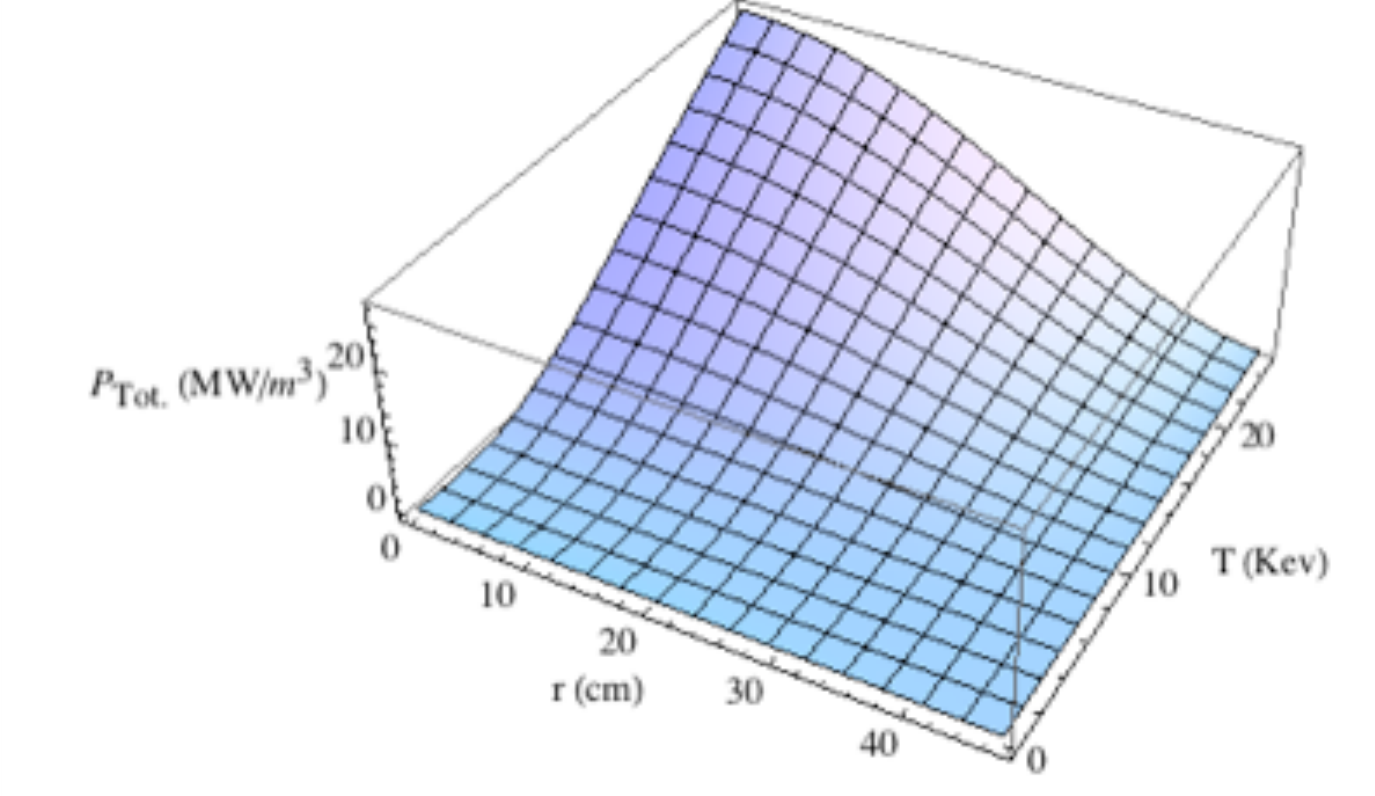}
}
\caption{ 
{\it Total source profile versus the minor radius and temperature in absence of additional source} ($P_{ICRH}\simeq 0$).
}
\label{fig3} 
\end{center}
  \end{minipage}
  \hfill
  \begin{minipage}[t]{.36\textwidth}
    \begin{center}
\hspace{-3.5cm}
\resizebox{1.5\textwidth}{!}{%
\includegraphics{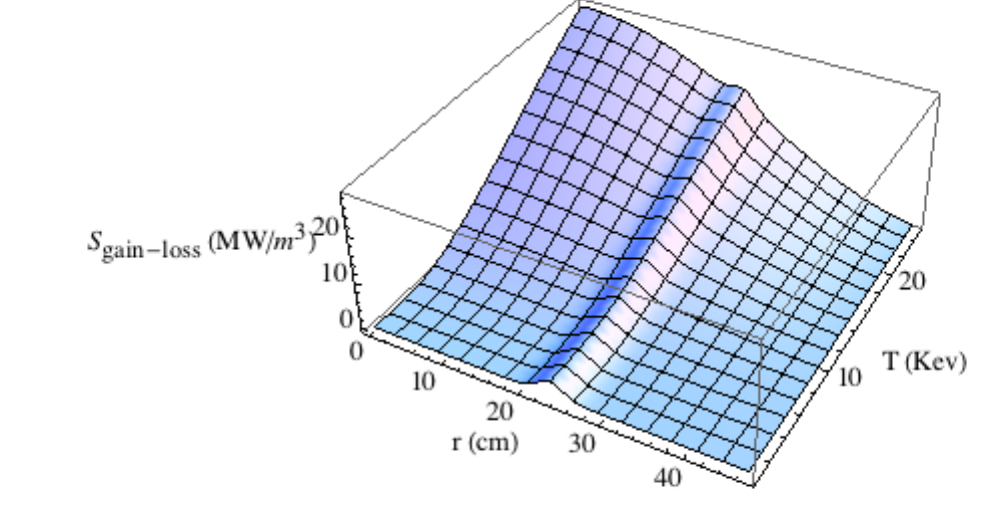}
}
\caption{
{\it Total source profile versus the minor radius and temperature in presence of ICRH} ($P_{ICRH}\simeq 1.5MW$).
}
\label{fig4}
\end{center}
  \end{minipage}
  \hfill
\end{figure*}
\noindent Finally, Eq.~({\ref{tl4}) results to be a highly non-linear second order ordinary differential equation in the radial variable $r$, submitted to the boundary conditions for the equilibrium temperature. The equilibrium temperature has been obtained by solving Eq.~(\ref{tl4}) numerically with the following conditions 
\begin{equation}\label{tl5}
{\rm and}\qquad d_rT\mid_{r=0}=0\qquad T\mid_{r=a}=pedestal = 20eV
\end{equation}
\noindent The first condition derives from the symmetry $T(r)=T(-r)$ close to the center of the Tokamak (we assume that, at $r=0$, the derivative of the temperature exists and does not diverge), and the choice of edge temperature $T=20eV$ is reasonable for several types of Ignitor L-mode plasmas. Figures~(\ref{fig5}) and (\ref{fig6}) show the temperature profiles, against the minor radius $r$, without RF power ($P_{ICRH}\simeq 0$), and when the power ICRH is injected into the plasma ($P_{ICRH}\simeq 1.5MW$), respectively.
\begin{figure*}
\hfill
  \begin{minipage}[t]{.48\textwidth}
    \begin{center}  
\hspace{-0cm}
\resizebox{1.1\textwidth}{!}{%
\includegraphics{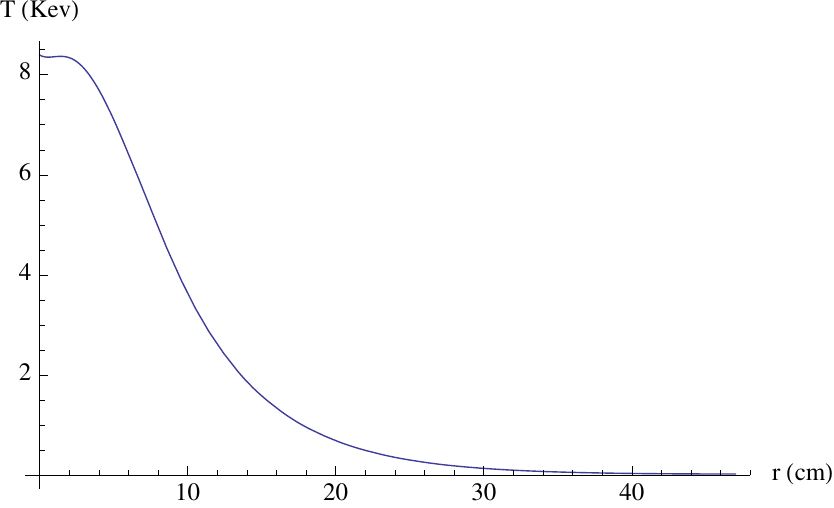}
}
\caption{
{\it Equilibrium Temperature profile when no power RF is provided. This solution has been obtained by solving numerically the steady state energy balance equation, with} ($P_{ICRH}\simeq 0$), {\it submitted to the boundary conditions Eqs~(\ref{tl5}).}
}
\label{fig5} 
\end{center}
  \end{minipage}
  \hfill
  \begin{minipage}[t]{.48\textwidth}
    \begin{center}
\hspace{-0cm}
\resizebox{1.1\textwidth}{!}{%
\includegraphics{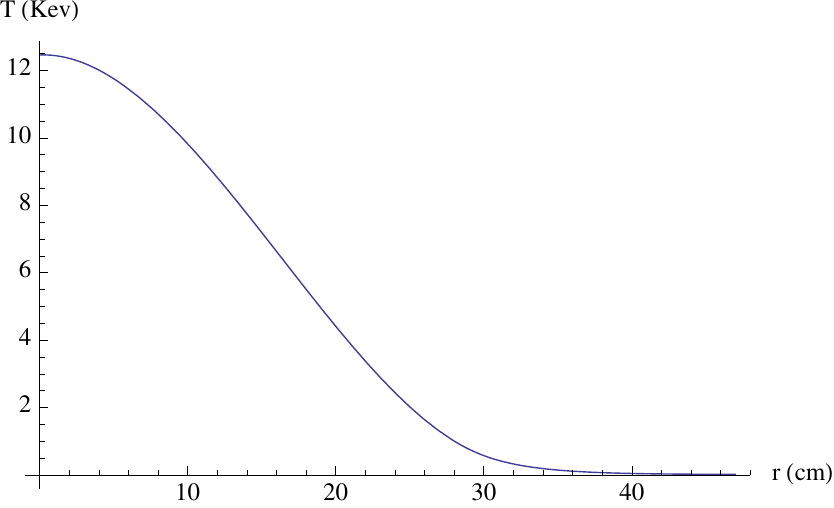}
}
\caption{
{\it Equilibrium Temperature profile when ICRH power is injected in the plasma} ($P_{ICRH}\simeq 1.5MW$). {\it This solution has been obtained by solving numerically the steady state energy balance equation, submitted to the boundary conditions Eqs~(\ref{tl5}).}
}
\label{fig6}
\end{center}
  \end{minipage}
  \hfill
\end{figure*}
\noindent Figures (\ref{fig7}) and (\ref{fig8}) illustrate the source profiles against the normalized radius $\rho$ for temperatures when no power RF is provided and when the power ICRH is injected into the plasma, respectively. These profiles have been obtained by inserting the equilibrium temperatures into the r.h.s. of Eq.~({\ref{tl4}), when $Q_{add}=0$ and $Q_{add}\neq 0$, respectively.
\begin{figure*}
\hfill
  \begin{minipage}[t]{.48\textwidth}
    \begin{center}  
\hspace{-0cm}
\resizebox{1.1\textwidth}{!}{%
\includegraphics{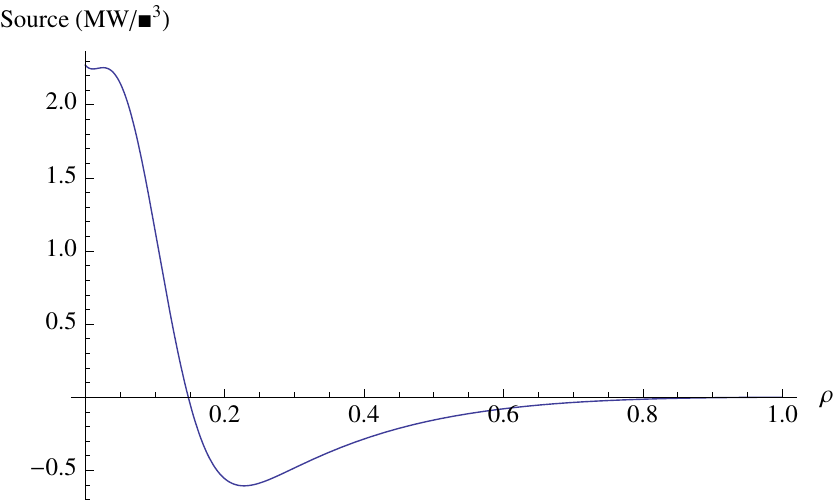}
}
\caption{
{\it Total source profile versus the normalized minor radius when} ($P_{ICRH}\simeq 0$).
}
\label{fig7} 
\end{center}
  \end{minipage}
  \hfill
  \begin{minipage}[t]{.48\textwidth}
    \begin{center}
\hspace{-0cm}
\resizebox{1.1\textwidth}{!}{%
\includegraphics{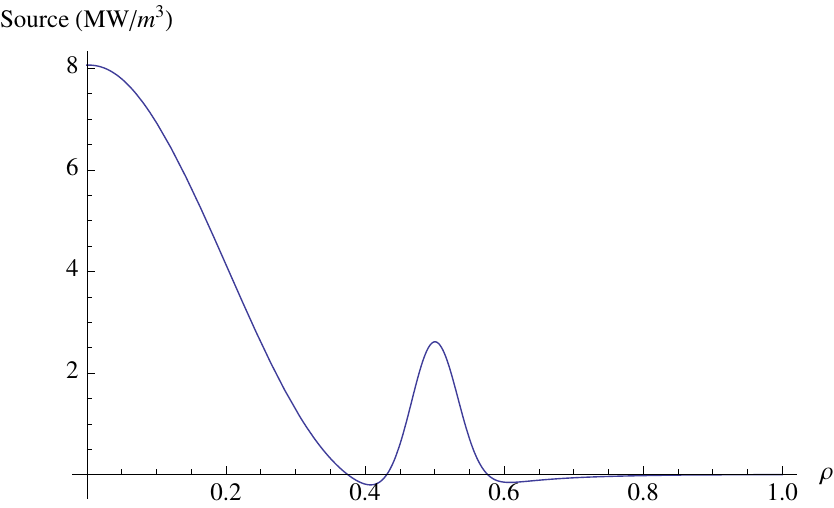}
}
\caption{
{\it Total source profile versus the normalized minor radius when} ($P_{ICRH}\simeq 1.5MW$).
}
\label{fig8}
\end{center}
  \end{minipage}
  \hfill
\end{figure*}
\noindent Figs.~(\ref{fig5}) (\ref{fig6}) (\ref{fig7}), and (\ref{fig8}), identify the region of the plasma where the profiles are not negligible (here referred to as the {\it core of the plasma}). This region ranges from $0\leq r\leq r_0$, with $r_0\simeq33,10(cm)$ and $r_0\simeq 25,14(cm)$ in absence and in presence of ICRH (with $P_{ICRH}\simeq1.5MW$) respectively. Note that $r_0$ is solution of the equation $T(r_0)={\bar T}\equiv V^{-1}\int T dV=(2/a^2)\int_0^axT(x)dx$. In the core of the Tokamak, the average temperature ${\bar T}_{core}$, defined as ${\bar T}_{core} =V_{core}\int_{core}T dV=(2/r_0^2)\int_0^{r_0}xT(x) dx$, are ${\bar T}_{core} =1.85KeV$ and ${\bar T}_{core} =6.14KeV$, for $P_{ICRH}=0$ and $P_{ICRH}\simeq 1.5MW$, respectively. Hence, in absence of ICRH, the average temperature in the core of the plasma does not reach the desired ignition temperature, which, as known, should be ${\bar T}_{core} > 4.4 KeV$. The additional ICRH power significantly increases the equilibrium temperature in the core of the Tokamak allowing the plasma to reach the ideal ignition temperature.

\section{Calculation of the stationary energy confinement time}\label{ect}

\noindent As known, the energy confinement time is defined as ratio between the total thermal energy $W_e$ in the plasma over the total energy rate through the boundary $\Gamma_E\mid_{boundary}$ {\it i.e.},
\begin{equation}\label{etc1}
\tau_E=\frac{W_e}{\Gamma_E\mid_{boundary}}
\end{equation}
\noindent with $W_e =3V^{-1}\int nT dV$ and $\Gamma_E\mid_{boundary}=V^{-1}\int\nabla\cdot{\bf q}_{tot} dV$. In our opinion, the most convenient way to estimate the energy confinement time correctly is to follow the procedure indicated by Freidberg \cite{freidberg}. The calculation begins by recalling the steady state $0-D$ plasma power balance relation. For this, let us reconsider the stationary power balance relation, Eq.~({\ref{etc2})
\begin{equation}\label{etc2}
\frac{n_0^2}{4}<\sigma v>_{D-T}E_\alpha-\frac{1}{4}c_Bn_0^2T^{1/2}+Q_{add}=\nabla\cdot{\bf q}_{tot}
\end{equation}
\noindent where we assumed that the particle density is constant, $n=n_0 = const.$, and $c_B$ , denotes the Bremsstrahlung constant [see Eq.~(\ref{s10})]. The $0-D$ steady state equation is obtained by assuming that the temperature profile is constant across the plasma cross section with magnitude equal to its average value ${\bar T}=V^{-1}\int T dV$. We get\footnote{Note that, for magnetically confined plasma in toroidal geometry, in the classical regime, the thermal diffusion strongly varies with the inverse of temperature. In particular, within the domain $0\leq r\leq a$, the vector field is not of class $C^1$ everywhere (see FIG. $2$). As known, under this circumstance, the divergence theorem does not apply. When the vector field is not of class $C^1$, we should proceed with the direct estimation of the integral of the divergence of the field over the volume: $P_Q=\int dV\nabla\cdot{\bf q}_{tot}=\int dV (P_Q+P_b+P_{ICRH})\neq 0$. In our problem, the correct way to perform calculations when ${\bf q}_{tot}=-<\kappa>dT/dr$ is not of class $C^1$ everywhere, is to proceed as follows: 1) Take into account the balance equation, Eq.~(\ref{etc2}), and 2) Follow the procedure indicated by Freidberg {\it i.e.}, to develop the integral by retaining only the leading order, according to the equations: $f(T)\approx f({\bar T})+f_1({\bar T)}\delta T+O(\delta T^2)$. Hence, $V^{-1}\int f(T)dV\approx f({\bar T})+O(\delta T)$. By following the above-mentioned procedure, we arrive at Eq.~(\ref{etc4}), which should be correct.}
\begin{eqnarray}\label{etc3}
\!\!\!\!\!\!\!\!\!\!\!\!\!\!\!\!\Gamma_E\mid_{boundary}=\Gamma_E\mid_{r=r_0} =\!V^{-1}\!\!\int\!\nabla\cdot{\bf q}dV\!=\!&&\!\!\!\frac{n_0^2}{4}E_\alpha<\sigma v>_{D-T}({\bar T})\!-\!\frac{1}{4}c_Bn_0^2{\bar T}^{1/2}\!+\!Q_{add}^0\nonumber\\
&&=\frac{W_e}{\tau_E}=\frac{3n_0{\bar T}}{\tau_E}
\end{eqnarray}
\noindent where Eq.~(\ref{etc1}) has been used for the definition of the energy confinement time, and $n_0\equiv n(r)$ and $Q^0_{add}\equiv Q_{add}(r_0)$ , respectively (recall that $r_0= 25.14cm$ and $T (r_0)={\bar T}$). Hence,
\begin{equation}\label{etc4}
\tau_E=\frac{12n_0{\bar T}}{E_\alpha n_0^2<\sigma v>_{D-T}({\bar T})-c_Bn_0^2{\bar T}^{1/2}+4Q_{add}^0}
\end{equation}
\noindent Finally, in presence of ICRH, with ($P_{ICRH}\simeq 1.5MW$), the estimated energy confinement time is $\tau_E\simeq 0.43sec$. It should be stressed that the analysis is underestimating the diffusive losses and this result should be regarded as just lower limits to the energy loss and hence an upper limit to the energy confinement time. Fig.~(\ref{fig9}) shows the profiles of $p\tau_E$ against the minor radius, in presence of ICRH, with ($P_{ICRH}\simeq 1.5MW$).
\begin{figure}\resizebox{0.45\textwidth}{!}{%
\includegraphics{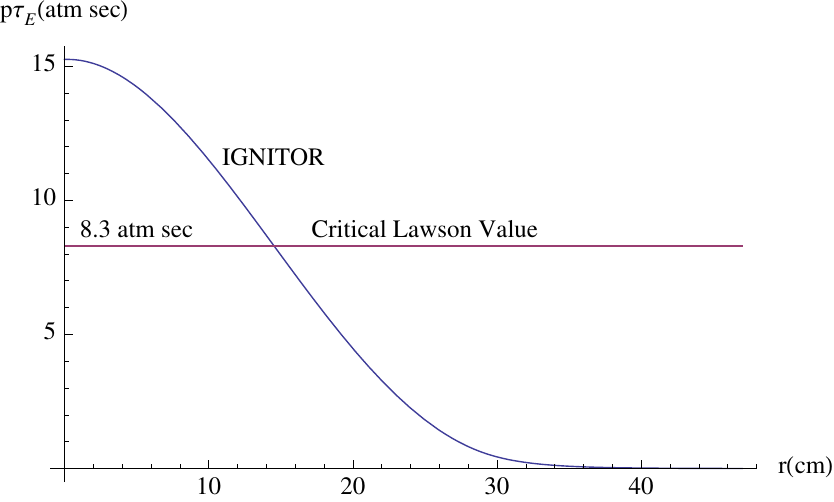}}
\caption{ \label{fig9} 
{\it The Lawson variable,} $p\tau_E$, {\it against the normalized minor radius for IGNITOR-plasma subject to ICRH. At the core of the plasma, the minimum values of} $T$ and $p\tau_E$, {\it required to satisfy the Lawson criterion for ignition, i.e.,} $T\geq 4.4 KeV$ {\it and} $p\tau_E\geq 8.3\ atm\ sec$, {\it are attained (see also Fig. (\ref{fig6}).}
}
\end{figure}
\noindent In conclusion, the additional heating is required during the startup transient phase in order to heat the plasma from its low initial temperature to the desired ignition temperature.

\section{Analysis of the thermonuclear instability}\label{ti}

\noindent The study of the stability of the solution, based directly on Eq.~(\ref{s6}), is quite complex. In the Appendix, we report a simplified analysis of stability where calculations are greatly simplified by eliminating the appearance of $<\gamma_r^e>$ and by assuming that the profiles (except temperature) are flat. These approximations will enable to determine the modes unstable. In this section we shall proceed in an even simpler way. We shall analyze the stability of the solution in the core of the plasma through the time-dependent form of $0-D$ power balance equation and by exploiting the (approximately) uniformity, in the core of the Tokamak, of the density profile. This approach will provide with the desired indications on thermal stability of the solution in the core of the plasma.

\noindent The space-time plasma dynamical equation, expressing the conservation energy relation for IGNITOR plasma in the standard model, can be brought into the form
\begin{equation}\label{ti1}
3\frac{\partial}{\partial t}(nT)=Q_\alpha+Q_b+Q_{add}-Q_\kappa
\end{equation}
\noindent where
\begin{equation}\label{tinon}
Q_\kappa\equiv\frac{1}{r}\frac{\partial}{\partial r}\bigl[r(<q_e>+<q_D>+<q_T>)\bigr]\nonumber
\end{equation}
\noindent By averaging Eq.~(\ref{ti1}) over space, we obtain the corresponding time dependent $0-D$ power balance equation:
\begin{equation}\label{ti2}
3n_0\frac{d}{dt}T=S_\alpha+S_b+S_{add}-S_\kappa
\end{equation}
\noindent where $S_\xi\equiv V_{core}^{-1}\int_{core}dV Q_\xi$ (with $\xi=\alpha,b,add,\kappa$) and we have assumed that $n=n_{core}=n_0\simeq const.$ The goal is now to examine the time dependance of a small perturbation $\delta T(t)$ of the equilibrium temperature ({\it i.e.}, $T(t)={\bar T}_{core} +\delta T(t)$ with $\delta T(t)/{\bar T}_{core} <<1$). In line with the Freidberg assumptions, we consider $n=n_0=const.$ and $S_{add}$ is a fixed quantity independent of temperature (i.e., $dS_{add}/dT=0$) \cite{freidberg}. At the leading order, a small perturbation $\delta T(t)$ satisfies the evolution equation :
\begin{equation}\label{ti3add1}
\frac{d\delta T}{dt}=\frac{1}{12}E_\alpha n_0\Bigl(\frac{d}{dT}<\sigma v>_{D-T}-{\tilde c}_bT^{-1/2}-\frac{12}{E_\alpha\tau_E}\Bigr){\Bigr\arrowvert}_{T={\bar T}_{core}}\!\!\!\!\delta T
\end{equation}
\noindent where ${\tilde c}_B\equiv2c_B/E_\alpha$. The stability condition can be further simplified by considering that (at linear order) the critical eigenvalue should be estimated {\it at equilibrium}. To this end, we recall that {\it at the equilibrium temperature} $S_\kappa=S_\alpha+S_{add}$ and $S_{add}=[(1-f_\alpha)/f_\alpha]S_\alpha$. Hence, we find $S_\kappa=S_\alpha/f_\alpha$, with $f_\alpha$ denoting the fraction $f_\alpha$ of the total heating power {\it i.e.}, $f_\alpha =S_\alpha /(S_\alpha +S_{add})$ \cite{freidberg} (so, $f_\alpha=1$ corresponds to ignition and $f_\alpha =0$ to no $\alpha$-power \footnote{Note that the definition for ignition $f_\alpha=S_\alpha/(S_\alpha+S_{add})$ is satisfied for $S_{add}=0$ at any temperature and nonzero fusion power $S_\alpha$. It only makes sense when the temperature exceeds some limit, such as the ideal ignition temperature $T\geq 4KeV$, so $S=S_b$.}). Now, by taking into account that $S_\kappa{\arrowvert}_{T={\bar T}_{core}}=n_0{\bar T}_{core}/\tau_E$ and $S_\alpha= n_0 E_\alpha<\sigma v>_{D-T}/12$, we get $12/(E_\alpha\tau_E)=<\sigma v>_{D-T}/(f_\alpha{\bar{T}_{core}})$ and Eq.~(\ref{ti3add1}) finally simplifies to 
\begin{equation}\label{ti3}
\frac{d\delta T}{dt}=\frac{1}{12}E_\alpha n_0\Bigl(\frac{d}{dT}<\sigma v>_{D-T}-{\tilde c}_bT^{-1/2}-\frac{<\sigma v>_{D-T}}{Tf_\alpha}\Bigr){\Bigr\arrowvert}_{T={\bar T}_{core}}\!\!\!\!\delta T
\end{equation}
\noindent Hence, the solution is stable if $\lambda({\bar T}_{core}, f_\alpha)\ < 0$ and unstable if $\lambda({\bar T}_{core}, f_\alpha)\ > 0$, where
\begin{equation}\label{ti4}
\lambda({\bar T}_{core}, f_\alpha)\equiv\frac{d}{dT}<\sigma v>_{D-T}{\Bigr\arrowvert}_{T={\bar T}_{core}}\!\!\!\!-{\tilde c}_B{\bar T}_{core}^{1/2}-\frac{1}{f_\alpha}{\bar T}^{-1}_{core}<\sigma v>_{D-T}{\Bigr\arrowvert}_{T={\bar T}_{core}}
\end{equation}
\noindent Fig.~(\ref{fig10}) reports on the generic profile of the critical eigenvalue $\lambda$ against the average temperature (in our case, ${\bar T}_{core}$) at the ignition value $f_\alpha =1$. Fig.~(\ref{fig11}) shows the critical eigenvalue $\lambda$ against the fraction of the total heating power, $f_\alpha$, estimated at ${\bar T}_{core} = 6.14KeV$. The dashed lines refer to the critical eigenvalue estimated by neglecting the Bremsstrahlung radiation.
\begin{figure*}
\hfill
  \begin{minipage}[t]{.5\textwidth}
    \begin{center}  
\hspace{-0cm}
\resizebox{1.1\textwidth}{!}{%
\includegraphics{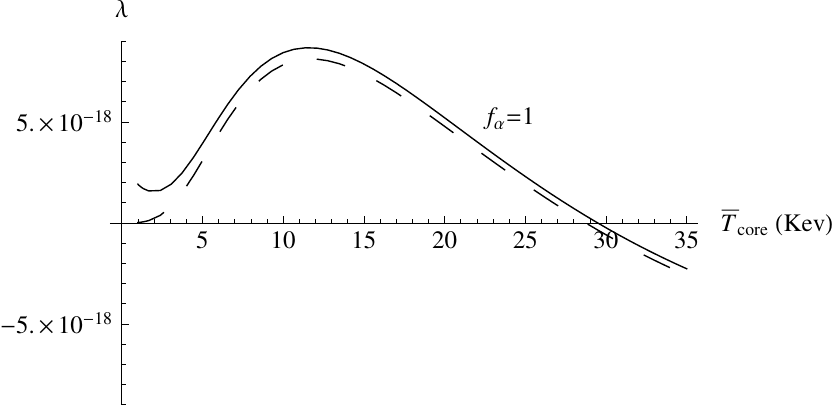}
}
\caption{
{\it Generic behavior of the critical eigenvalue} $\lambda$ {\it versus the average temperature (in our case,} ${\bar T}_{core}${\it), at the ignition value} $f_\alpha =1$. {\it The dashed line corresponds to} $\lambda$ {\it profile estimated by neglecting the Bremsstrahlung radiation.}
}
\label{fig10} 
\end{center}
  \end{minipage}
  \hfill
  \begin{minipage}[t]{.43\textwidth}
    \begin{center}
\hspace{-0cm}
\resizebox{1.1\textwidth}{!}{%
\includegraphics{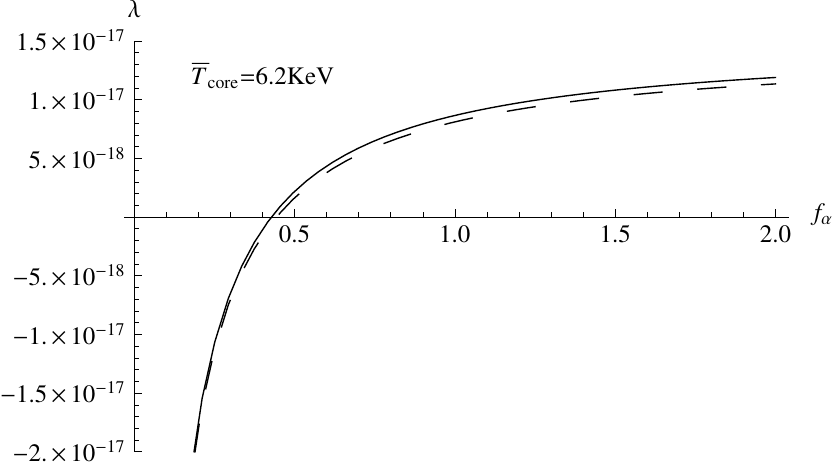}
}
\caption{
{\it Critical eigenvalue} $\lambda$ {\it against} $f_\alpha$, {\it at} ${\bar T}_{core}=6.2keV$. {\it The dashed line corresponds to the estimation of} $\lambda$ {\it by neglecting the Bremsstrahlung effect}.
}
\label{fig11}
\end{center}
  \end{minipage}
  \hfill
\end{figure*}
\noindent In line with our expectations, at ${\bar T}_{core} = 6.14KeV$ and $f_\alpha =1$ (ignition), the core of the plasma is unstable. The Bremsstrahlung effect provides a negligible contribution.

\section{ Conclusions}\label{conclusions}
\noindent One of the objectives of this work is to study in detail the equilibrium, and stability properties, of the temperature evolution in burning fusion IGNITOR-plasma, in presence of ICRH. Although in this respect many manuscripts already appeared in literature, our aim is to determine the equilibrium temperature and to study the stability of the solution, by making use of the plasma dynamical equations and transport relations, which are rigorously obtained by kinetic theory. In addition, our approach gives some new insights concerning the thermal runaway problem and, in particular, the relation between the $0-D$ and $1-D$ models. Here a scenario is considered where IGNITOR is led to operate in a slightly sub-critical regime by adding a small fraction of  ${}^3He$ to the nominal $50$$\%$-$50$$\%$ Deuterium-Tritium mixture. At the first step, we considered the simplest case where the transport coefficients are determined by kinetic theory applied to classical plasma in a toroidal geometry. The obtained results may be sketched as follows.
\begin{description}
\item { {\bf i)} We determined the temperature equilibrium profile solely by kinetic theory i.e., without the {\it auxilium} of ad hoc models for the transport coefficients};
\item { {\bf ii)} We showed that in the core of the plasma the thermal solution is unstable, and we estimated that the value of the confinement time is $\tau_E\simeq 0.43sec.$};
\item{ {\bf iii)} The additional heating,ICRH, is required during the start up transient phase in order to heat the plasma from its low initial temperature to the desired ignition temperature};
\item{{\bf iv)} We showed that the ICRH heating in the IGNITOR experiment is expected to trigger the thermonuclear burning by means of the RF coupled power. The use of the ICRH can be switched {\it on} and {\it off} along with the plasma parameter evolution and in particular with the temperature. If we apply ICRH to a plasma, characterized by a subcritical ignition regime, we have shown that it is possible to trigger a thermal instability, by switching off the ICRH the regime can be recast to a subcritical one. This means that in the subcritical regime the difference between power lost and alpha heating is compensated by additional ICRH heating, which should be able to increase the global plasma temperature via collisions between ${}^3He$ minority and the background $D-T$ ions}.
\end{description}

\noindent It is well known that a realistic estimation of energy confinement time should account the turbulent contributions and, in particular, the strong anomalous diffusion in the outer plasma. However, this is a very complex task. Our analysis is obviously valid in the core of the plasma, corresponding to the region $0\leq r\leq r_0$ with $T(r_0)={\bar T}$. The clear identification of the core of the plasma is due to our choice of boundary conditions, {\it i.e.}, a pedestal temperature with a thermal derivative that is zero at the edge. In the core the presence of the auxiliary ICRH heating is responsible of the triggering of the instability.

\noindent We mention another aspect concerning the instability problem. Here, we have considered an IGNITOR (type) device, characterized by a large $B$-field and small dimensions. Of course, these conditions enormously simplified calculations, since a large-$B$ field tends {\it to freeze} turbulent effects. In IGNITOR, indeed, we have evaluated that the non-linear (turbulent) contribution to the transport is not dramatic owing to the fact that IGNITOR operates with a very strong external magnetic field. The magnetic field has a stabilizing effect on the turbulence. Evidence of this fact can be deduced by the calculation done by means of the TFT code \cite{sonnino1} where an evaluation of the strength of the non linear contribution has been established for the electron and ion fluxes. The result is that the difference is sufficiently weak and the linear theory can be used safely. In addition, the peculiarity of IGNITOR is that, since this reactor works at sufficiently low temperature (positive slope of temperature curve), the instability can develop as soon as the criteria of ignition are met. Another argument is that in IGNITOR the collisionality regime is essentially {\it banana} for most of the discharge radius being of the Pfirsch-Schl$\ddot{\rm u}$ter type only in a small portion at the edge and at the center. However we decided to dedicate a publication {\it per se} to study in a deeper manner all the transport regimes of IGNITOR by covering the various radial zones. Other Reactor tokamak designs, based on low-$B$ field and large dimensions have also been analyzed, but only in terms of heuristic experimental scalings. This obviously does not hold for Tokamak Reactor like ITER or DEMO, where the turbulence can play a crucial role in the determination of transport coefficient. The main difference with respect to IGNITOR is that DEMO (which is characterized by low magnetic field, large dimensions, and very high temperature) is far from developing a thermal instability. DEMO in fact is characterized by a negative slope of the temperature curve and for this reason is thermally stable \cite{freidberg}. In addition, for these reactors, a realistic estimation should take into account the strong anomalous diffusion in the outer plasma and, under this conditions, the temperature profiles estimated by using the classical thermal diffusion would probably result in highly unrealistic shapes. Anyhow, a deeper analysis in which a comparison between both approaches (high field and low temperature) and low field and large dimension will be performed more extensively in a dedicated work.  

\noindent We would like also clarify another crucial point: the role of accumulation of reaction ashes ${}^4He$, which may eventually quench the thermonuclear process. In reality, in this work we considered the emergence and development of the thermal instability just at the end of the flattop. In this scenario, the presence of the alpha particle is still too low to give some evaluable effect on the dynamic of the reaction. Obviously, during the flattop, the presence of a consistent fraction of ${}^4He$ could induce the quench of the thermonuclear reaction below the useful threshold. In fact the ashes play the same role of the impurities by unbalancing the {\it good} ratio of the reactant (50$\%$ Deuterium and 50$\%$ Tritium). Also in this case, in our idea, the ICRH power turns to be a useful tool in giving a boost at the plasma temperature to compensate the presence of the impurities that are degrading the reaction rate.

\noindent Now, we should proceed step-by-step. In the next step we shall consider the general situation, where the transport coefficients are determined by considering all the collisional transport regimes ({\i.e.}, the  classical, Pfirsch-Schl$\ddot{\rm u}$ter and banana transport regimes), and the nonlinear contributions are no longer neglected. The results will be compared successively with the solutions obtained by using a turbulent transport model, like the gyro-Bohm model. All of this will be subject of future works.

\section{Acknowledgements}
One of us (GS) is indebted to Gy$\ddot{\rm o}$rgy Steinbrecher, of the University of Craiova (Romania), for the fruitful discussions concerning the topic presented in the Appendix. GS is also very grateful to Alberto Sonnino, of the Karlsruhe Institute of Technology (KIT) - Germany and the Universit{\' e} Catholique de Louvain (UCL) - Belgium, for his assistance in performing numerical calculations.

\appendix
\noindent\section{Determination of the Modes Unstable - Simplified Calculations.}\label{appx}

In this Section, we report a quite simplified analysis of the stability of the equilibrium temperature, showing the methodology allowing the determination of the modes unstable. A semi-quantitatively accurate approximation is to assume that all the profiles, except temperature, and the coefficients are flat. This is not a very good approximation because, actually, the profiles and the coefficients (in particular) are not flat, but the approximation greatly simplified the analysis. Hence, this Appendix should be understood only as an example of calculation with a view to illustrating the procedure.

\noindent Let us put $T(r,t) = T_0(r)+\delta T(r,t)$ where $T_0(r)$ is the equilibrium temperature, solution of Eq.~({\ref{etc2}), and $\delta T(r,t)$ the temperature perturbation. From Eqs~(\ref{s6}), (\ref{tl1}) and (\ref{etc2}), and taking into account the expression of $S_{gain-loss}$, we find the evolution equation for the perturbation $\delta T(r,t)$
\begin{eqnarray}\label{appx1}
\!\!\!\!\!\!\!\!3n_0\frac{\partial\delta T}{\partial t}-<\kappa_{tot}>{\Bigr\arrowvert}_{T=T_0(r)}\frac{\partial^2\delta T}{\partial r^2}=&&
\!\!\!\!-\frac{n_0^2}{4}\Bigl(2c_BT_0(r)^{-1/2}-E_\alpha\frac{\partial}{\partial T}<\sigma v>{\Bigr\arrowvert}_{T=T_0(r)}\\
&&-\frac{4}{n_0^2}\frac{\partial^2T_0}{\partial r^2}\frac{\partial}{\partial T}<\kappa_{tot}>{\Bigr\arrowvert}_{T=T_0(r)}\Bigr)\delta T\nonumber\end{eqnarray}
\noindent where we have taken into account that $\partial_TQ_{add}= 0$ and we have neglected terms higher than the first order in $\delta T$. In addition, we have supposed that the contribution $r^{-1}(\partial_rT)\partial_r(r<\kappa_{tot}>)$ may be neglected with respect to $<\kappa_{tot}>\frac{\partial^2T}{\partial r^2}$. We assume now that the thermal conductivity $<\kappa_{tot}>$ coefficient and $-\frac{n_0^2}{4}\bigl(2c_BT_0(r)^{-1/2}\!\!-E_\alpha\partial_T<\sigma v>\mid_{T=T_0(r)}\bigr)$ are constant and estimated at the average temperature $T={\bar T}$, {\it i.e.}
\begin{eqnarray}\label{appx2}
&&\!\!\!\!\!\!\!\!\!\!\!\!\!\!\!\!<\kappa_{tot}>{\Bigr\arrowvert}_{T=T(r_0)} = <\kappa_{tot}>{\Bigr\arrowvert}_{T={\bar T}}=const.\nonumber\\
&&\!\!\!\!\!\!\!\!\!\!\!\!\!\!\!\!-\frac{n_0^2}{4}\Bigl(2c_BT_0(r)^{-1/2}-E_\alpha\frac{\partial}{\partial T}<\sigma v>{\Bigr\arrowvert}_{T=T_0(r)}\Bigr)=\\
&&\qquad\qquad\qquad\qquad\qquad\quad-\frac{n_0^2}{4}\Bigl(2c_B{\bar T}^{-1/2}-E_\alpha\frac{\partial}{\partial T} <\sigma v>{\Bigr\arrowvert}_{T={\bar T}}\Bigr)=const.\nonumber
\end{eqnarray}
\noindent In order to have an idea on the validity of first approximation in Eq.~({\ref{appx2}), we report in Figs~(\ref{fig12}) and (\ref{fig13}) the total average thermal coefficient $<\kappa_{tot}>$ versus the normalized minor radius $\rho$, in absence and in presence of ICRH, respectively. These profiles have been obtained by putting the equilibrium temperature-profiles, given in Figs~(\ref{fig5}) and (\ref{fig6}), into $<\kappa_{tot}>_{CL}$, respectively.
\begin{figure*}
\hfill
  \begin{minipage}[t]{.43\textwidth}
    \begin{center}  
\hspace{-0cm}
\resizebox{1.1\textwidth}{!}{%
\includegraphics{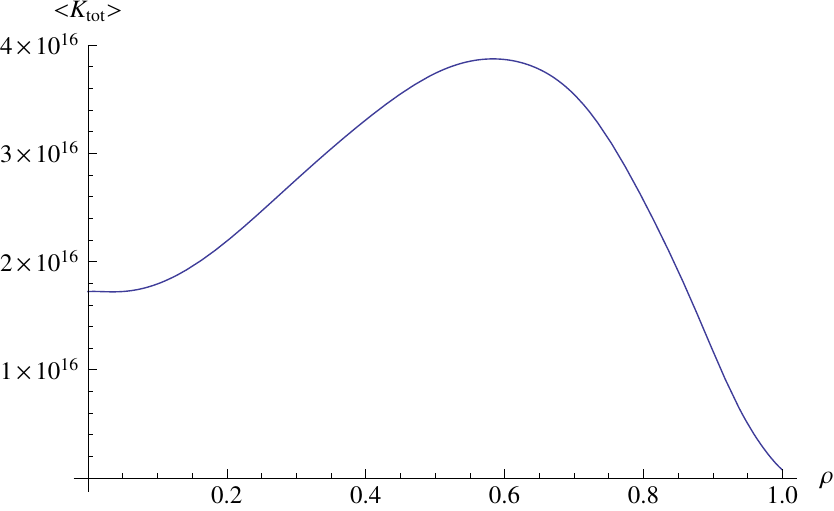}
}
\caption{
{\it Total average thermal conductivity (measured $cm^{-1} sec^{-1}$), against the normalized minor radius, in absence of additional sources. This profile has been obtained by putting the equilibrium temperature given in Fig.~(\ref{fig5}) into $<\kappa_{tot}>$.}
}
\label{fig12} 
\end{center}
  \end{minipage}
  \hfill
  \begin{minipage}[t]{.43\textwidth}
    \begin{center}
\hspace{-0cm}
\resizebox{1.1\textwidth}{!}{%
\includegraphics{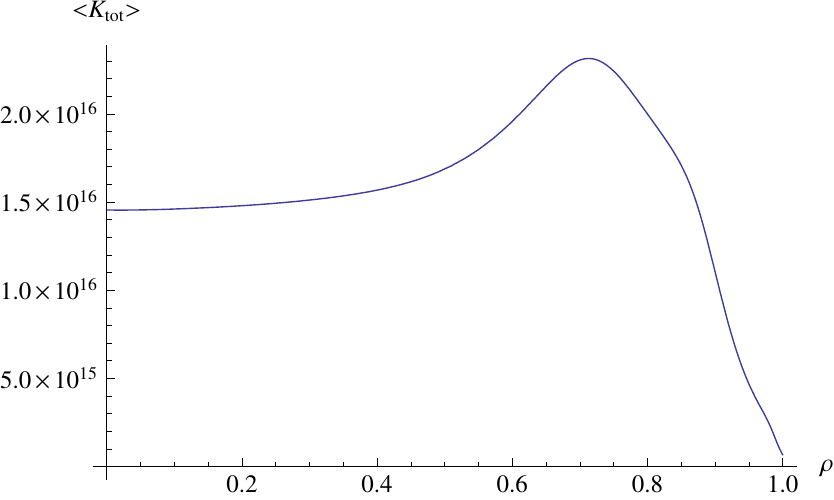}
}
\caption{
{\it Total average thermal conductivity (measured $cm^{-1} sec^{-1}$), against the normalized minor radius, in presence of ICRH, with ($P_{ICRH}\simeq1.5MW$). This profile has been obtained by putting the equilibrium temperature given in Fig.~(\ref{fig6}) into $<\kappa_{tot}>$.}
}
\label{fig13}
\end{center}
  \end{minipage}
  \hfill
\end{figure*}
\noindent Hence, Eq.~(\ref{appx1}) takes the form
\begin{equation}\label{appx3}
\frac{\partial}{\partial t}\delta T =D\frac{\partial^2}{\partial r^2}\delta T+\beta\delta T
\end{equation}
\noindent with
\begin{equation}\label{appx4}
D\equiv\frac{<\kappa_{tot}>\mid_{T={\bar T}}}{3n_0}\quad {\rm and}\quad \beta =-\frac{n_0}{12}\Bigl(2c_B{\bar T}^{-1/2}\!\!-E_\alpha\frac{\partial}{\partial T}<\sigma v>{\Bigr\arrowvert}_{T={\bar T}}\Bigr)
\end{equation}
\noindent The boundary conditions may be determined by imposing that both temperature and its derivative do not fluctuate at the boundary. So, we have to solve Eq.~(\ref{appx1}) subject to
\begin{equation}\label{appx5}
\frac{d}{dr}\delta T{\Bigr\arrowvert}_{r=0}\!\!\!\!=0\qquad{\rm and}\qquad \delta T{\Bigr\arrowvert}_{r=a}=0
\end{equation}
\noindent By setting $\delta T(r,t)=e^{-\omega t}f(r)$, we get
\begin{equation}\label{appx6}
D\frac{d^2f(r)}{dr^2}+(\beta+\omega)f(r)=0
\end{equation}
\noindent with $d_rf\!\!\!\mid_{r=0}\ =f\!\!\!\mid_{r=a}=0$. The solution of Eq.~(\ref{appx6}) can be brought into the form $f(r)=\sum_{k=0}^n{\hat f}_k\cos(k r)$. We find
\begin{eqnarray}\label{appx7}
{\hat f}_0&&=0\qquad({\rm for}\ \ k=0)\nonumber\\
-Dk^2+(\beta+\omega)&&=0\qquad ({\rm for}\ \ k\neq 0)
\end{eqnarray}
\noindent The boundary conditions (\ref{appx5}) provide the relation between $\omega$ and the modes $n$. Indeed,
\begin{equation}\label{appx8}
\cos(ka)=0\qquad\Longrightarrow\quad ka=\frac{\pi}{2}+n\pi\ \ \ (n=\pm1,\pm2,\cdots)
\end{equation}
\noindent By substituting Eq.~(\ref{appx8}) into Eq.~(\ref{appx6}), we get
\begin{equation}\label{appx9}
\omega(k)=-\beta+Dk^2\qquad\Rightarrow\qquad \omega(n)=-\beta+D\Bigl(\frac{\pi}{a}\Bigr)^2\Bigl(n+\frac{1}{2}\Bigr)^2
\end{equation}
\noindent By taking into account Eq.~(\ref{appx4}), we find that the {\it modes unstable} satisfy the inequality
\begin{equation}\label{appx10}
\Bigl(n+\frac{1}{2}\Bigr)^2\frac{a^2n_0^2}{4\pi^2}\frac{1}{<\kappa_{tot}>\mid_{T={\bar T}}}\Bigl(E_\alpha\frac{\partial}{\partial T}<\sigma v>{\Bigr\arrowvert}_{T={\bar T}}\!\!\!\!-2c_B{\bar T}^{-1/2}\Bigr)
\end{equation}
\noindent In particular, the Goldstone mode ($n=0$) is unstable if
\begin{equation}\label{appx11}
a^2n_0^2E_\alpha\frac{\partial}{\partial T}<\sigma v>{\Bigr\arrowvert}_{T={\bar T}}\!\!\!\!-2a^2n_0^2c_B{\bar T}^{-1/2}\!\!-\pi^2<\kappa_{tot}>{\Bigr\arrowvert}_{T={\bar T}}\!\!>0
\end{equation}
\noindent 
A more refined calculation has been proposed in ref.~\cite{coppi1} where the relevant mode involving the growth of the electron temperature perturbations is tridimensional and radially localized around a given rational magnetic surface. Clearly, the onset and evolution of this kind of {\it ribbon} modes have to be considered in order to envision and predict how a condition of global ignition can be reached \cite{coppi2}.

\bigskip
\end{document}